\documentclass[showpacs,twocolumn]{revtex4-1}

\usepackage{graphicx}
\usepackage[fleqn]{amsmath}
\usepackage{verbatim}
\usepackage{bbm} 
\usepackage{eucal}
\usepackage{amsthm}
\usepackage{amsfonts}
\usepackage{color}

\newcommand{\be}{\begin{eqnarray}}
\newcommand{\ee}{\end{eqnarray}}
 \newcommand{\bra}{\langle } 
 \newcommand{\ket}{\rangle } 
 \newcommand{\id}{1\!\!1}

\newtheorem{definition}{Definition}

\usepackage{amsthm}

\begin{document}
\title{The Security of Quantum Key Distribution using a Simplified Trusted Relay}
\author{William Stacey,$^1$ Razieh Annabestani,$^1$ Xiongfeng Ma,$^2$ Norbert L\"{u}tkenhaus$^1$}
\affiliation{$^1$ Institute for Quantum Computing, University of Waterloo, Waterloo, Ontario, Canada, N2L~3G1}
\affiliation{$^2$ Center for Quantum Information, Institute for Interdisciplinary Information Sciences, Tsinghua University, Beijing, China}

\begin{abstract}

We propose a QKD protocol for trusted node relays. Our protocol shifts the communication and computational weight of classical post-processing to the end users by reassigning the roles of error correction and privacy amplification, while leaving the exchange of quantum signals untouched. We perform a security analysis for this protocol based on the BB84 protocol on the level of infinite key formulas, taking into account weak coherent implementations involving decoy analysis. 

\end{abstract}

\date{\today} 
\maketitle


\section{introduction}\label{s:intro}

Quantum key distribution (QKD) \cite{Scarani09, BB84} is one of the immediate applications of quantum information theory. However, QKD still faces several technical  hurdles. One  challenge is to implement long-distance QKD. For point-to-point QKD, in which one user sends quantum signals directly to another user, the key rate is approximately bounded by the single-photon transmittance of the channel \cite{Takeoka13}. For fibre-optic implementations, this leads to an exponential reduction of the key rate, resulting in unattractive key rates for distances over a few hundred kilometres --- even when considering optimistic system performance. The maximum distance is typically limited by the dark count rate of the detectors, leading to a vanishing key rate at distances around 200-300 km. In the long term, advanced quantum repeaters \cite{Briegel98,Muralidharan2013} promise practical long distance QKD; however, they are currently under development on a fundamental research level. Trusted relays offer a short-term solution. They have already been demonstrated in several active QKD networks \cite{Elliott05,Peev09,Sasaki11,Stucki11}, and have been proposed for use in satellite QKD nodes \cite{Meyer11,Yin13,Toyoshima11}. 

 In standard trusted relays, full QKD protocols are executed between nearest neighbours in a series of trusted nodes. Each node publicly announces the parity of the two keys it holds, enabling the end users to create a shared key. It is important that all nodes are trusted, as each node could reproduce the final key. One drawback of this approach is that each intermediate node is involved in full QKD protocols with its nearest neighbours, including post-processing steps such as error correction and privacy amplification. These protocol elements can be demanding in terms of computational resources and the communication bandwidth. This can lead to problems, for example, when using lightweight satellites which are restricted in computation and communication. Even for unrestricted relays, the total computation and communication overhead of a large chain of trusted nodes is significant.

In this paper, we present an alternate version of the trusted relay, which reduces the requirements of the intermediate nodes by shifting post-processing tasks to the end users \cite{LutkenhausPatent13}.  We will refer to this variant as the simplified trusted relay (STR). Other simplifications to the trusted relay have been suggested, such as delayed privacy amplification \cite{Fung12}. 

STRs are similar to trusted relays; however, instead of each node making announcements to connect the completed secret keys that it shares with its neighbours, each node makes announcements based on the raw data that it generates in the quantum phase of the QKD protocol. The end users then carry out the remainder of the post-processing based on their raw data and the announcements from the trusted relay. This reduces the required complexity of each node, as well as the computational load on the trusted nodes. 

It is important to note that in both the standard and the simplified trusted relay, the intermediate nodes must be completely trusted. This trust assumption can be reduced using independent paths in connection with secret-sharing protocol ideas \cite{Beals08,Salvail10}. Additionally, one could use encryption of public announcements to reduce the impact of compromised nodes that satisfy the typical {\em honest-but-curious} constraints. However, the basic structure of trusted relays demands a minimal level of trust in the intermediate nodes. 

In Section \ref{s:proto}, we describe the general STR protocol. We further list detailed steps for a specific realization of the STR protocol which employs the quantum phase of the BB84 protocol \cite{BB84}. In Section \ref{s:keyrate}, we examine the security of this STR protocol and derive a key rate formula for the ideal case where the legitimate parties exchange qubit signal states. The security proof is then extended to optical modes in Section \ref{s:decoy}.


\section{STR Protocol}\label{s:proto}
 
 \begin{figure}
\begin{center}
\includegraphics[scale=0.77]{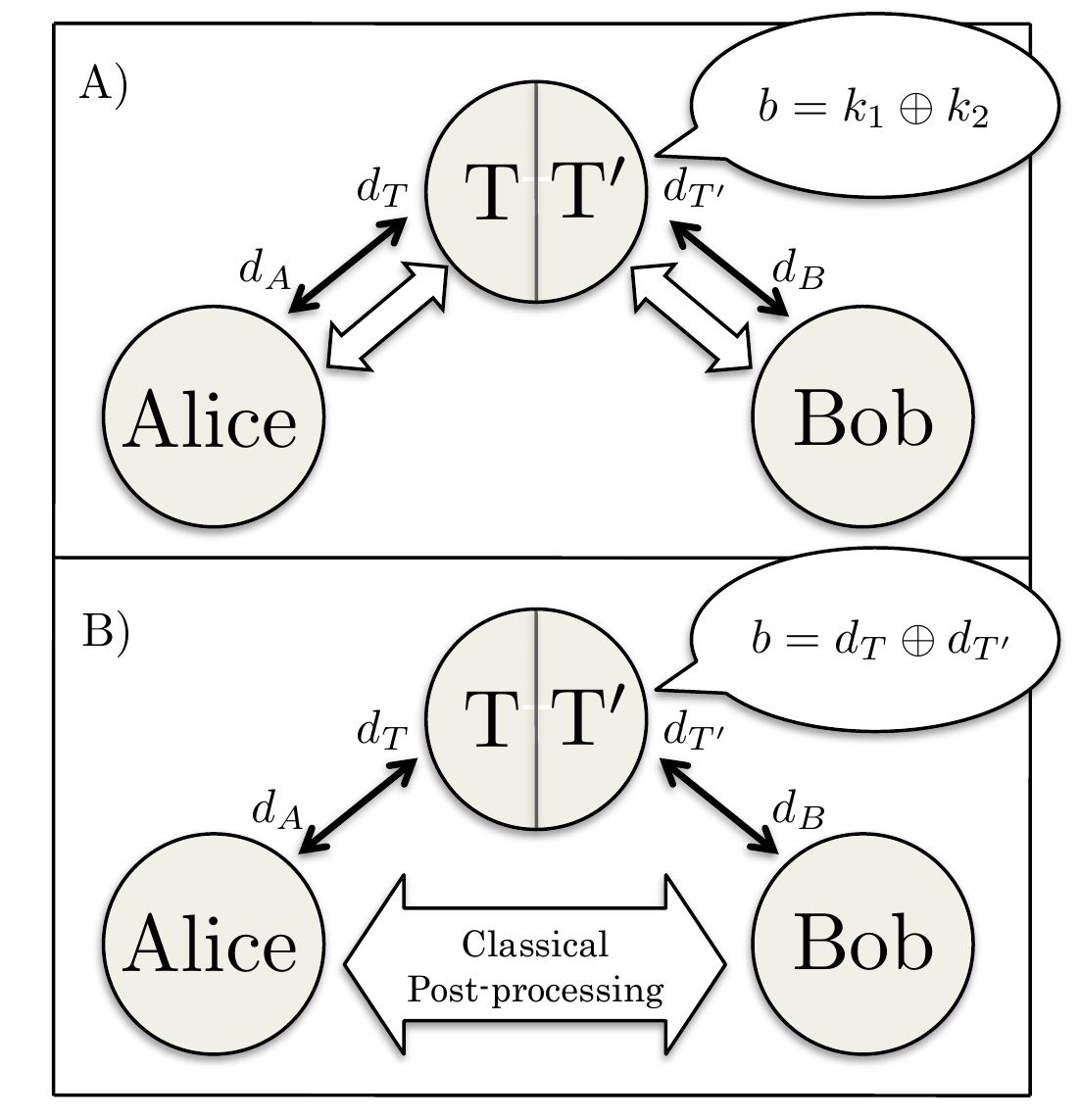} 
\caption{(A) A conventional trusted relay with a single node. Quantum signals (black arrows) are used to establish raw data ($d$) in each link. Secret keys ($k_1$, $k_2$) are distilled from the raw data, using classical post-processing (white arrows). Bob recovers Alice's key, using the parity announcement, i.e. $k_2 \oplus b = k_1$. Here $\oplus$ denotes bitwise modulo-2 addition. (B) A single node STR. Before error correction and privacy amplification, Alice, Bob, and the node (T,T$^\prime$) hold the respective raws keys, $d_A$, $d_B$, $d_{T}$, and $d_{T^\prime}$. The trusted node announces the parity ($b$) of the raw keys that it holds. In the absence of errors $d_{A}=d_{T}$, $d_{B}=d_{T^\prime}$ and $d_{B}\oplus b = d_{A}$. Alice and Bob carry out the majority of post-processing. All parties share authenticated classical channels.
}
\label{schematic.pdf}
\end{center}
\end{figure}

STR protocols closely resemble conventional trusted QKD relays. In each link, quantum signals are distributed and measured to derive a set of measurement results and settings. In a conventional trusted  relay, neighbouring parties would then use classical post-processing to create secret keys in each link. Each node would connect the two secret keys it shares with its neighbours by announcing the bitwise parity of these keys. However, as we will show, a trusted relay may securely function with less assistance from the nodes. For STRs, neighbouring nodes only need to perform the quantum stage of a QKD protocol. By sending and measuring quantum signals, the legitimate parties generate raw data of quantum origin. The nodes connect the raw data by a public parity announcement, analogous to conventional trusted relays. The completion of the QKD protocol, involving error correction and privacy amplification, is left entirely to the end users (Alice and Bob). For a single node, the STR protocol is contrasted to a conventional trusted relay in Figure \ref{schematic.pdf}.

 In this paper, we demonstrate the basic idea of STRs by focussing on a particular STR protocol based on the BB84 QKD protocol. Note that many variations of this protocol exist for which our analysis will directly apply. Moreover, our analysis can be generalized in a straightforward manner to other protocols, such as the 6-state protocol \cite{6-state,Bechmann99} or continuous-variable protocols \cite{Grosshans02,Silberhorn02}.
 
The BB84-based protocol proceeds as follows: 

\begin{enumerate}

\item \textbf{Point-to-Point Data Creation.} Alice and Bob and all intermediate nodes perform this step with their nearest neighbours.
\begin{enumerate}
\item \textbf{State creation and distribution:} Alice chooses a basis $u_A \in \{Z,X\}$ with probability $p_{u_A}$, then selects a bit value $x \in \{0,1\}$ with uniform probability and prepares the corresponding BB84 qubit state $|\phi ^{u_A} _x\ket$. Alice records the state she created, then sends the signal to the nearest trusted node. Alice carries out this process $N$ times, where $N$ is suitably large. Similarly, each node prepares signals and sends them to the next node in the relay. The last node sends signals to Bob. 
\item \textbf{Measurement:} Each node (and Bob) locally select a basis $u \in \{Z,X\}$ with probability $p_{u}$ and perform a projective measurement in that basis, denoted by the positive operator valued measure (POVM) $\textbf{M}^{u}$. To distinguish between the two roles each node plays, we use $u_{T_j}$ to denote the $j$-th node's measurement basis, while $u_{T^\prime_j}$ denotes the $j$-th node's state preparation basis. (For the case of a single node, we will disregard the index.) The choice of basis in each link is independent, i.e. $u_{T_j}$ is independent of $u_{T^\prime_j}$. Bob and the nodes record their measurement outcome, as well as the basis in which they measured. 
\item \textbf{Sifting:} The legitimate parties reveal their measurement and preparation bases. In each link, data are kept only when an event was detected and the basis choices coincided. All other data are discarded, reducing the size of the data strings from $N$ to $n$. In the following sections, we will use $u_1 := u_A = u_{T_1}$ to denote the basis choice for an event that has survived sifting in the first link. Similarly, the basis in the $i$-th link is denoted by $u_i$. For each signal in each link, Alice and Bob record $u_i$.
\item \textbf{Keymap:} The legitimate parties map their remaining data into classical bit strings (raw keys, $d$), by mapping the BB84 states they have sent and/or measured into raw bits using the rule $|\phi ^{u}_x\ket \rightarrow ``x"$, where $x \in\{0,1\}$. Alice now holds the raw key $d_A = \{x_1, \dots, x_n\}$. For clarity, we denote Bob's raw key as $d_B = \{y_1, \dots ,y_n \}$, where $y$ denotes Bob's measurement outcome. Similarly, the $j$-th node holds the raw keys $d_{T_j} = \{t_{j,1}, \dots, t_{j,n}\}$ and $d_{T^\prime_j} = \{t^\prime_{j,1}, \dots, t^\prime_{j,n}\}$.
\end{enumerate}

\item \textbf {Node Announcement.} Only the intermediate nodes perform this step.
\begin{enumerate}
\item \textbf{Parity announcement:}  Each node announces the bitwise parity of the two raw keys that it holds. For the $j$-th node, $\textbf{b}_j = \{b_{j,1}, \dots , b_{j,n} \}$, where $b_{j,i} = t_{j,i} \oplus t^\prime_{j,i} $.
\end{enumerate}

\item \textbf{End user protocol.} Only the end users Alice and Bob perform this step.
\begin{enumerate}
\item \textbf{Processing of Parity Announcements} Bob adds each parity announcement to his own raw key to obtain a new raw key $d^\prime_B := \{y_1^\prime, \dots ,y_n^\prime \}$, where $y_i^\prime =(\bigoplus_j b_{j,i})\oplus y_i$. In the absence of errors, $d^\prime_B = d_A$ and Bob recovers Alice's raw key.
\item \textbf{Parameter estimation:} Alice and Bob determine the error rate for each basis combination, including the basis choices of the intermediate nodes. For $m$ nodes, Alice and Bob therefore determine $2^{m+1}$ error rates. If the error is suitably low, they continue the protocol. Otherwise, they abort.
\item \textbf{Error correction and Privacy Amplification:} If the protocol is not aborted, Alice and Bob carry out one-way error correction and privacy amplification to obtain secure keys.
\end{enumerate}

\end{enumerate}

In order to extend our security analysis from collective to general attacks, we will later require that the protocol has \emph{permutation invariant} properties (see Section \ref{ss:catga}). We assume that both error correction and privacy amplification are carried out in a permutation invariant manner. While permutation invariant methods exist, most practical methods of error correction and privacy amplification typically are not permutation invariant. It remains an open question whether this condition is truly necessary.

For the following theoretical analysis, the prepare-and-measure scenario in the above protocol can equivalently be described as a source-replacement scheme \cite{Bennett92}. In this thought set-up, each source generates the entangled state $|\Phi \ket \in \mathcal{H}_{SS^\prime}$, where $|\Phi \ket = (1/\sqrt{2})(|00\ket + |11\ket) $. Instead of sending the signal state, the source then sends the second half of the entangled system ($S^\prime$). Using a POVM $\textbf{M}^{u}$ chosen with probability $p_u$, the legitimate party which has prepared the entangled state, performs a projective measurement on the system $S$, preparing $S^\prime$.

The source-replacement scheme reveals an important property. In the \emph{point-to-point data creation} step, the roles of signal state preparation and measurement may be interchanged in each link without affecting the basic security of the protocol. Similarly, one could use a scheme based on physical entanglement to perform this step, where both parties measure signals prepared by an untrusted intermediate source. One could also use a measurement-device independent QKD protocol \cite{Lo12} to establish raw data in each link by making use of additional untrusted nodes. Note that while the qubit security proof in Section \ref{s:keyrate} applies to all of these situations, the extension to realistic implementations in Section \ref{s:decoy} looks specifically at the protocol as detailed above. 


\section{Qubit Security Proof}\label{s:keyrate}
Using the formalism developed in Refs. \cite{Kraus05, Renner05, Renner05Thesis}, we establish a rate at which secret key bits may be derived from the raw key. We begin our analysis by examining the ideal case where qubit signal states are exchanged over lossless channels. Furthermore, we examine the asymptotic limit where the legitimate parties exchange a large number of signals. In this limit, we can consider an eavesdropper (Eve) to make collective attacks without loss of generality \cite{Christandl09}. (In this context, collective attacks are defined for each group of signals exchanged along the chain between Alice and Bob  that are matched by the parity announcements. Eve can attack all signals contributing to the group jointly. We will justify this definition in Section \ref{ss:catga}.)  Using this analysis as a foundation, we extend the security to include loss and multi-photon sources in Section \ref{s:decoy}. 

To simplify notation, let us first examine the case where our relay has one trusted node. The analysis for an arbitrary number of nodes follows similarly. After distributing signals, Alice, Bob, and the trusted node hold the tripartite state $\rho_{ATB}$. For simplicity, we use $T$ to denote the two qubit-spaces composing the node's system. In order to ensure that Eve is limited only by the laws of quantum mechanics, we allow Eve complete control over the purification $|\Psi\ket_{ATBE} $. Eve's choice of purification is limited only by the observed quantities:

\begin{definition} \label{def1}
Let $\Gamma$ be the set of all states $\rho_{ATBE}$ consistent with all observables measured by the legitimate parties.
\end{definition}

From Refs. \cite{Renner05Thesis, Kraus05, Renner05}, the rate at which secret key bits may be distilled from raw key bits (the key rate) is 

\begin{equation}\label{keytheoretical}
r = H(K_A)_{obs} - \text{leak}^{EC}_{obs} - \max \limits_{\Gamma} \chi(X:E)
\end{equation}
where $H(K_A)_{obs}$ is the observed Shannon entropy of Alice's raw key and leak$^{EC}_{obs}$ is the actual information leaked during error correction. Note that both these quantities can easily be calculated or bounded from observed data. The Holevo quantity between Alice and Eve is denoted by $\chi(X:E)$. Note that $E$ includes all classical communication available to Eve prior to error correction. We adopt the notation of Ref. \cite{Ferenczi12}, defining $\chi(\rho_{ATBE},\textbf{M}^{u_1}_A):= \chi(X:E)$, where $\textbf{M}^{u_1}_A$ represents Alice's POVM conditioned on her basis choice.


\subsection{Announcements and Postselection}

In the STR protocol, the legitimate parties in each link postselect on data where the basis choices ($u$) match. In order to do so, they publicly announce the basis in which states were prepared or measured. Any events in which the basis choice did not coincide are discarded. Additionally, the node announces the parity ($b$) of the raw keys that it holds. We must take into account how these announcements affect security.

The quantum treatment of postselection is represented by a trace preserving map, which takes $\rho_{ATBE}$ to $\bar{\rho}_{ATBE}$. The formalism behind this postselection method can be found in Ref. \cite{Ferenczi12}. The announcements are represented by a classical register $C$ held by the legitimate parties and Eve. The postselected state is therefore block-diagonal, given explicitly by

\begin{equation} \label{psstate}
\bar{\rho}_{ATBE} = \sum \limits_{u_1,u_2,b} p_{u_1}p_{u_2}  p_{ b|u_1,u_2} \rho ^{u_1,u_2, b} _{ATBE} \otimes |u_1 u_2 b\ket \bra u_1 u_2 b| _C. 
\end{equation}
Here $p_{u}$ represent the probability of each basis announcement and $p_{ b|u_1,u_2}$ is the conditional probability associated with the parity announcement. Furthermore, $\rho ^{u_1,u_2, b} _{ATBE}$ is the state held between all parties, conditioned on a given combination of announcements. Note that the choice of bases, and thus the corresponding  announcements, are determined by local randomness. As a result, the state conditioned on $u_1$ and $u_2$ alone is the same as the original state, i.e. $\rho ^{u_1,u_2} _{ATBE} = \rho _{ATBE}$. However, the parity announcement is determined by a measurement outcome; the effect proves to be non-trivial. Following the postselection formalism in Ref. \cite{Ferenczi12}, $\rho ^{u_1,u_2, b} _{ATBE} \neq \rho _{ATBE}$. 


\subsection{Parameter Estimation}

In order to derive an analytic key rate, we will relax the conditions on the shared state. Instead of using all observables, we look only at specific error rates. For each basis combination, the error rate between Alice's raw key, $d_A$, and Bob's corrected raw key, $d_{B}^\prime$, is given by 
\begin{equation}\label{errorrate}
e^{u_1,u_2} =  \sum \limits_b p_{b|u_1,u_2}  \sum \limits_{x\neq y\oplus b}tr_{AB}(M^{u_1}_{A,x} \otimes M^{u_2}_{B,y} \rho^{u_1,u_2, b}_{AB}) 
\end{equation}
where $M^{u_1}_{A,x}$ and $M^{u_2}_{B,y}$ denote Alice and Bob's POVM elements. With this in mind, let us define $\tilde{\Gamma}$, a superset of $\Gamma$:

\begin{definition} \label{def2}
Let $\tilde{\Gamma}$ be the set of all states $\rho_{ATBE}$ consistent with the set of basis-dependent error rates $e^{u_1,u_2}$.
\end{definition}

Given that $\Gamma \subseteq \tilde{\Gamma}$, we may lower-bound the key rate by replacing the maximization over $\Gamma$ in Eq. (\ref{keytheoretical}), with a maximization over $\tilde{\Gamma}$.


\subsection{Symmetries in the STR Protocol}

We begin the security proof by reducing the size of the set $\tilde{\Gamma}$. The optimal attack is shown to occur when the reduced state $\rho_{ATB}$ is diagonal with respect to a basis consisting of tensor products of Bell states (see Eq. (\ref{e:bellsymstate})). This is a direct result of symmetries in the BB84 signal states. In the following section, we will make use of the Bell-diagonal form of $\rho_{ATB}$.

As outlined in Ref. \cite{Ferenczi12}, the form of Eve's optimal attack can often be simplified if the following conditions are met:
\begin{enumerate}
	\item The composition of the Holevo quantity and any postselection mapping applied to $\rho_{ATBE}$ is concave. 
	\item A set of states $\beta = \{\rho^i_{ATBE}\}$ is found, where each state yields the same chosen observables as the original state $\rho_{ATBE}$.  In our case, these chosen observables are basis-dependent error rates.
	\item Each state $\rho^i_{ATBE}$ results in the same Holevo quantity as the original state $\rho_{ATBE}$.
\end{enumerate}
If these three conditions are met, it follows that the Holevo quantity is maximized by a state of the form $\frac{1}{|\beta|} \sum _i \rho^i_{ATBE}$.

In order to satisfy the first of the above conditions, we first use basic properties of the von Neumann entropy to rearrange the key rate in Eq. (\ref{keytheoretical}):
\begin{flalign}
&r \geq \sum \limits_{u_1,u_2} p_{u_1} p_{u_2} H(K_A^{u_1,u_2})_{obs} - \text{leak}^{EC}_{obs}&\label{key2}\\
& - \max \limits_{\tilde{\Gamma}} \sum \limits_{u_1,u_2} p_{u_1} p_{u_2}  \chi \left(\sum \limits_b p_{b|u_1,u_2} \rho_{ATBE}^{u_1,u_2,b} \otimes |b \ket \bra b|_C, \textbf{M}^{u_1}_A \right).\nonumber& 
\end{flalign}
Here $H(K_A^{u_1,u_2})_{obs}$ denotes the entropy of Alice's key data arising from each basis-combination. 

The concavity of the composition of the Holevo quantity and the mapping in Eq. (\ref{key2}) follows directly from the fact that the mapping is linear and the Holevo quantity is concave \cite{Ferenczi12}.
Drawing inspiration from Refs. \cite{Kraus05,Ferenczi12}, we apply correlated Pauli operations in each link to generate a set of states $\beta = \{\rho^i_{ATBE}\}$ (see Appendix \ref{appendix1}). In the appendix, we further show that the basis-dependent error rates and the Holevo quantity in Eq. (\ref{key2}) are invariant for each of these states. Therefore we restrict our search for Eve's optimal attack, to states of the form $\frac{1}{|\beta|} \sum _i \rho^i_{ATBE}$.

 In Appendix \ref{appendix2}, we show that the reduced form of the averaged state is diagonal with respect to a basis consisting of tensor products of Bell states. Therefore we restrict our search for Eve's optimal attack from $\tilde{\Gamma}$ to $\tilde{\Gamma}_\text{Bell}$, where $\tilde{\Gamma}_\text{Bell}$ is defined to be the set of all states that are consistent with the observed basis-dependent error rates, and also have the reduced form 
\begin{align}\label{e:bellsymstate}
\rho^\text{Bell}_{ATB} := \sum \limits_{a,b,a^\prime,b^\prime=0}^1\alpha_{a,b,a^\prime,b^\prime} &|\Phi_{a,b}\ket \bra \Phi_{a,b}|_{A,T}&\\
& \otimes |\Phi_{a^\prime,b^\prime}\ket \bra \Phi_{a^\prime,b^\prime}|_{T,B}.&\nonumber
\end{align}
Above $\alpha_{a,b,a^\prime,b^\prime}$ are an arbitrary set of normalized eigenvalues, while $ |\Phi_{a,b}\ket$ are the four Bell states:
\begin{equation}
|\Phi_{a,b}\ket := \frac{1}{\sqrt{2}}\sum \limits_{k=0}^1 (-1)^{ak}|k\oplus b\ket |k\ket.
\end{equation}
Here $\oplus$ denotes modulo-2 addition. 

\subsection{Qubit Key Rate}

The evaluation of the key rate makes use of similarities between the measurements in an STR protocol and Bell measurements. If the intermediate nodes perform joint Bell measurements on the entangled quantum systems that they share with their respective neighbours, then this corresponds to entanglement swapping, providing the end user with entangled states (as well as information about the relevant reference frame determined by the set of outcomes of the Bell measurements). In this case, the end users can establish a secret key based only on the observed correlations analyzed separately for each announced set of Bell measurements, without further involvement from the intermediate nodes, or even trust in the nodes.

The primary observation linking the entanglement swapping picture of quantum relays to the STR protocol is that a Bell measurement can be deconstructed into a parity and a phase bit measurement on two qubits. The parity result of this measurement, $b$, is identical to the parity announcement in our STR protocol (up to local Hadamard operations, as mentioned below). However, in the STR protocol, the phase measurement result, denoted by the bit $a$, is suppressed. Still, as we show below, we can use this relationship to evaluate the secret key rate of the STR protocol. Note that there are distinct differences between an entanglement-swapping quantum relay and the STR protocol. In the STR protocol we cannot prove security based on Alice and Bob's observations alone. Instead, we are required to trust the measurements that the intermediate nodes perform. The complexity of our analysis is increased by the fact that the actual form of the Bell measurements discussed above depends on the basis choices used in the two links. 

 In the STR protocol, the trusted node measures each link in either the $X$- or $Z$-basis and announces the parity bit of the measurement results. As stated above, we can imagine an alternate protocol where the node carries out a Bell measurement for the respective basis, followed by an announcement of the parity component of the Bell measurement result, but not the phase component. If we denote the $Z$-basis on the $i$-th qubit with $u_i=0$ and the $X$-basis with $u_i=1$, the rotated Bell basis is given by $\{ H^{u_1} \otimes H^{u_2} | \Phi_{a,b}\ket  \}_{a,b=0}^1$, where $H^{u_i}$ denotes a Hadamard matrix raised to the power $u_i$ and $|\Phi_{a,b}\ket$ are the four Bell states. The rotated Bell basis is explicitly 
\begin{align}\label{e:bellbasis}
&\{ H^{u_1} \otimes H^{u_2} | \Phi_{a,b}\ket  \}_{a,b=0}^1 =& \\
& \left\{
     \begin{array}{lr}
       \{|\Phi_{0,0}\ket,|\Phi_{0,1}\ket,|\Phi_{1,0}\ket,|\Phi_{1,1}\ket\} & : u_1 =0, u_2 =0\\
       \{|\Phi^\prime_{0,0}\ket,|\Phi^\prime_{0,1}\ket,|\Phi^\prime_{1,0}\ket,|\Phi^\prime_{1,1}\ket\} & : u_1 =0, u_2 =1 \\
       \{|\Phi^\prime_{0,0}\ket,|\Phi^\prime_{1,0}\ket,|\Phi^\prime_{0,1}\ket,|\Phi^\prime_{1,1}\ket\} & : u_1 =1, u_2 =0 \\
       \{|\Phi_{0,0}\ket,|\Phi_{1,0}\ket,|\Phi_{0,1}\ket,|\Phi_{1,1}\ket\} & : u_1 =1, u_2 =1 
     \end{array}
   \right.& \nonumber \\
&|\Phi^\prime_{a,b}\ket := (\id \otimes H) |\Phi_{a,b}\ket.&
\end{align}
When $u_1 = u_2$, the above set is simply a permutation of the Bell states. Similarly, when $u_1 \neq u_2$ the set is a permutation of the Bell states, up to a local unitary.

To simplify the security analysis, we may consider announcing the phase bit to Eve (but not to Alice and Bob), effectively putting a lower bound on the key rate of the alternative protocol. Given that the security of the alternative protocol is equivalent to the STR protocol, we therefore lowerbound the key rate of the STR protocol. Intuitively, leaking the phase bit to Eve cannot decrease Eve's knowledge of the key. This notion can be made rigorous using the monotonicity of the quantum relative entropy under partial trace \cite{Ruskai02}. This insight leads to the bound 
\begin{flalign} \label{box2}
 &\max \limits_{\tilde{\Gamma}_\text{Bell}}  \chi \left(\sum \limits_b p_{b|u_1,u_2} \rho_{ATBE}^{u_1,u_2,b} \otimes |b \ket \bra b|_C, \textbf{M}^{u_1}_A \right)&\nonumber\\
  \leq&\max \limits_{\tilde{\Gamma}_\text{Bell}}  \chi \left(\sum \limits_{a,b} p_{a,b|u_1,u_2} \rho_{ATBE}^{u_1,u_2,a,b} \otimes |a,b \ket \bra a,b|_C, \textbf{M}^{u_1}_A \right)&\\
 =  & \max \limits_{\tilde{\Gamma}_\text{Bell}} \Bigg(  \sum \limits_{a,b} p_{a,b|u_1,u_2}\chi (\rho^{u_1,u_2,a,b} _{ATBE}, \textbf{M}^{u_1}_A)& \nonumber\\
  &+ H(K_A^{u_1,u_2}) - \sum \limits_{a,b} p_{a,b|u_1,u_2} H(K_A^{u_1,u_2,a,b}) \Bigg). \nonumber&
\end{flalign}
Above, $p_{a,b|u_1,u_2}$ denotes the probability associated with the parity and phase measurement, conditioned on the measurement basis. Similarly, $\rho^{u_1,u_2,a,b} _{ATBE}$ is the joint state conditioned on the particular announcement combination and $K_A^{u_1,u_2,a,b}$ is the conditional key data. The above equality makes use of simple properties of the von Neumann entropy. Note for each state in $\tilde{\Gamma}_\text{Bell}$, it holds that $H(K_A^{u_1,u_2}) = H(K_A^{u_1,u_2,a,b})$ for all values of $u_1,u_2,a,b$. Therefore the second and third term in the maximization vanish.

A simple method for placing an upper bound on the above maximization is to maximize each term $\chi (\rho^{u_1,u_2,a,b} _{ATBE}, \textbf{M}^{u_1}_A)$ individually. If the maximization specifies $Z$- and $X$-error rates (or equivalent restrictions) for each conditional state $\rho_{ATBE}^{u_1,u_2,a,b}$, the result of the maximization is well known \cite{Scarani09}. However, upon inspection, it is not immediately apparent that $\tilde{\Gamma}_\text{Bell}$ contains suitable restrictions. First, the observed error rates arise from the conditional states $\rho_{ATBE}^{u_1,u_2,b}$, not $\rho_{ATBE}^{u_1,u_2,a,b}$. Second, it is not clear that  the $Z$- and $X$-error rates can be \emph{simultaneously} determined for each conditioned state (see Eq. (\ref{errorrate})). We address the first concern by considering the hypothetical error rates $e^{u_1,u_2,a,b}$ and later invoking the concavity of the binary entropy to derive a key rate dependent only on the observed error rates $e^{u_1,u_2}$. The second concern is addressed using a relation among the conditioned states (see Eq. (\ref{there})).

To address the concerns mentioned above, let us write the set $\tilde{\Gamma}_\text{Bell}$ in terms of the hypothetical error rates for each Bell announcement, i.e. $\tilde{\Gamma}_\text{Bell}$ is the set of all states $\rho^\text{Bell}_{ATBE}$ consistent with the error rates $e^{u_1,u_2,a,b}$, such that $\sum_{a,b} p_{a,b|u_1,u_2} e^{u_1,u_2,a,b}$ equals the observed error rates $e^{u_1,u_2}$. With these two conditions in mind, we define:

\begin{definition}
Let $\tilde{\Gamma}_{\text{hidden}}$ be the set of all states $\rho^\text{Bell}_{ATBE}$ consistent with all error rates $e^{u_1,u_2,a,b}$.
\end{definition}

\begin{definition}
Let $S_{\text{obs}}$ be the set of all error rates $e^{u_1,u_2,a,b}$ such that $\sum_{a,b} p_{a,b|u_1,u_2} e^{u_1,u_2,a,b}=e^{u_1,u_2}$.
\end{definition}
The maximization in Eq. (\ref{box2}) can then be treated as two separate maximizations:
\begin{flalign} 
&  \max \limits_{\tilde{\Gamma}_\text{Bell}} \chi (\rho^{u_1,u_2,a,b} _{ATBE}, \textbf{M}^{u_1}_A)  =   \max \limits_{S_{\text{obs}}}  \max \limits_{\tilde{\Gamma}_{\text{hidden}}} \chi (\rho^{u_1,u_2,a,b} _{ATBE}, \textbf{M}^{u_1}_A) . &
\end{flalign}
 For this approach to be useful, we must first show that the set $\tilde{\Gamma}_{\text{hidden}}$ contains suitable restrictions on each state $\rho^{u_1,u_2,a,b}_{ATBE}$. The form of the Bell measurement reveals that certain sets of the conditioned states $\rho^{u_1,u_2,a,b}_{ATBE}$ are related by trivial relabelings. Explicitly,
\begin{flalign}\label{there}
&\rho^{u_1=i,u_2=j,a=k,b=l}_{ATBE}=\rho^{u_1=i\oplus 1,u_2=j \oplus1,a=l,b=k}_{ATBE} \ \forall i,j,k,l.&
\end{flalign}
The above relations allow us to derive $X$- and $Z$-basis error rates for each conditioned state. For example, the $X$-error rate arising from the conditioned state $\rho^{u_1=0,u_2=0,a=k,b=l}_{ATBE}$ is given by $e^{u_1=1,u_2=1,a=l,b=k}$.

We can now maximize each term, $\chi (\rho^{u_1,u_2,a,b} _{ATBE}, \textbf{M}^{u_1}_A)$, as if it had arisen from an independent protocol (after making use of the above relation). Given that we are only interested in an upper bound, we can choose to only examine relevant restrictions when maximizing each term. This maximization can now be handled using techniques outlined in Appendix A of Ref. \cite{Scarani09}:
\be \label{punchline}
   \max \limits_{\tilde{\Gamma}_{\text{hidden}}}\chi (\rho^{u_1=i,u_2=j,a=k,b=l}_{ATBE}, \textbf{M}^{u_1}_A)&\\
   \leq h(e^{u_1=i\oplus 1,u_2=j \oplus1,a=l,b=k})& \quad \forall i,j,k,l. \nonumber
 \ee
Although the individual error rates, $e^{u_1, u_2,a,b}$, are unknown, we can arrive at a useful key rate by first using the fact that the conditional probability $p_{a,b|u_1,u_2}$ respects similar relations to Eq. (\ref{there}),
\be
p_{a=i,b=j|u_1=k,u_2=l} = p_{a=j,b=i|u_1=k\oplus1,u_2=l\oplus1}.
\ee
Then, by using the concavity of the binary entropy, with consideration of Eq. (\ref{keytheoretical}) and Eq. (\ref{box2}), we find the key rate to be
\be\label{keyrate}
 r &\geq& \sum \limits_{u_1,u_2} p_{u_1} p_{u_2} H(K_A^{u_1,u_2})_{obs}- \text{leak}^{EC}_{obs}\\
  &&-  \sum \limits_{u_1,u_2}  p_{ u_1\oplus 1}p_{u_2\oplus 1} h(e^{u_1, u_2}). \nonumber 
\ee
Note that after using the concavity of the binary entropy, the maximization over $S_{\text{obs}}$ is trivial, as each element in $S_{\text{obs}}$ results in the same key rate.

The same analysis may easily be extended to the case where $m$ trusted nodes are used. In this case, we describe the basis choices for the $m+1$ links are described as $\textbf{u} = \{ u_1, \ldots ,u_{m+1}\}.$ If the legitimate parties share the postselected state $\bar{\rho}_{AT_mBE}$, and if we define the set $\tilde{\Gamma}_m$ similarly to Def. \ref{def1}, then we may apply the same analysis to find
\be
\max \limits_{\tilde{\Gamma}_m} \chi (\bar{\rho}_{AT_mBE},\textbf{M}^{u_1}_A) \leq
  \sum_{\substack{u_1, \ldots ,u_{m+1} }}
 p_{\tilde{\textbf{u}}}  h(e^{\textbf{u}}).
\ee
where the vector $\tilde{\textbf{u}} =  \{ u_1\oplus 1, \ldots ,u_{m+1}\oplus 1\}$. Analogous to Eq. (\ref{keyrate}), $p_{\tilde{\textbf{u}}}$ is the probability of the announcement combination $\tilde{\textbf{u}}$. Similarly, $e^{\textbf{u}}$ is the rate of errors between Alice and Bob conditioned on both \textbf{u}.

 \begin{figure}
\begin{center}
\includegraphics[scale=0.8]{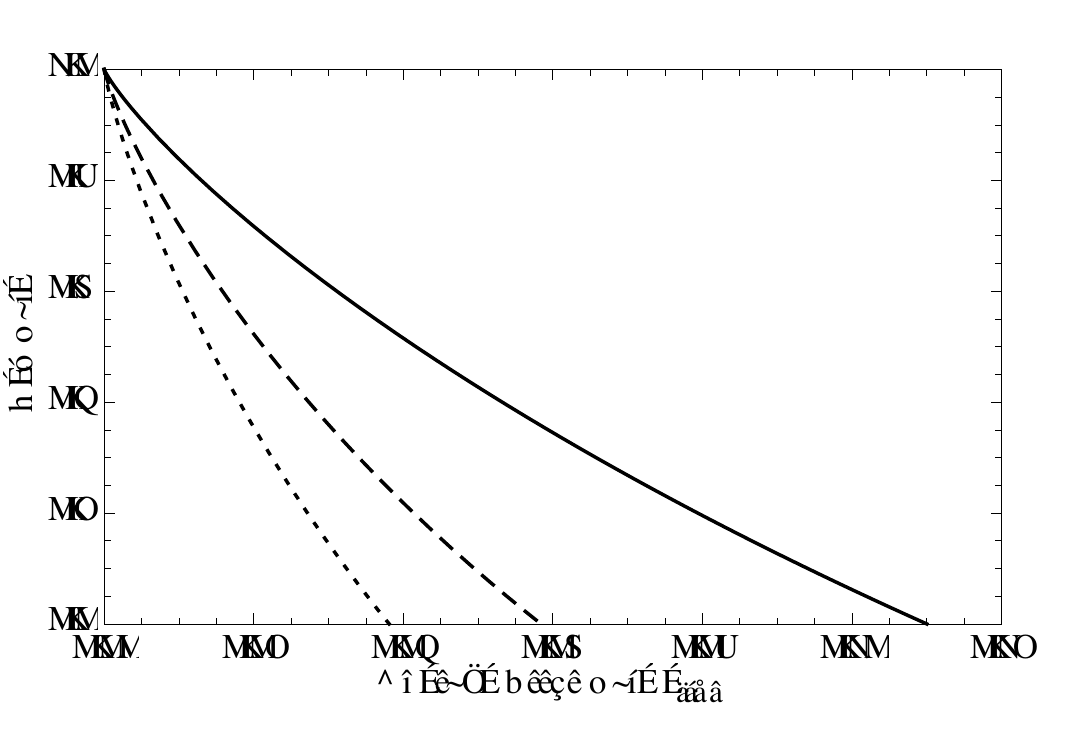} 
\caption{The rate at which key bits can be derived from the raw key, in the limit of an infinite key length, for a conventional trusted node (solid line), an STR with one node (dashed line), and an STR with two nodes (dotted line), as a function of the error rate in each link, $e_\text{link}$. The plot assumes realistic error patterns. Error correction is assumed to be done in the Shannon limit.
}
\label{fig_rate_error.pdf}
\end{center}
\end{figure}

The qubit security of an STR differs from conventional trusted relays in two major ways. First, the key rate is only a function of the error rate between Alice and Bob. Second, the addition of new nodes in an STR protocol degrades the key rate due to compounding errors. This prevents STRs from being extended to arbitrary distances. The qubit key rates for one and two nodes are plotted in Fig. \ref{fig_rate_error.pdf}, along with the key rate for a conventional trusted relay. As shown, the simplicity of the STR protocol comes at the cost of a reduced key rate. Note that this does not take into account computational time; for particular scenarios, the \emph{key generation rate per unit time} may in fact be higher for STRs.

Allowing one of the intermediate nodes to generate error-correcting information will reduce the amount of information revealed to Eve during error correction. In a possible variant of the above STR protocol, one of the intermediate nodes may define the key map. That node generates one-way error correcting information and broadcasts it to Alice and Bob. While our previous analysis \emph{does not} apply to this situation, a canonical calculation of the key rate shows that the security of this alternate protocol is nearly equivalent to the above protocol --- if a single node is used and the bases are chosen with uniform probability. As shown in Ref. \cite{Staceythesis}, the resulting key rate for this \textit{node-focused} case is 
\begin{flalign}
&r_\text{node-focused}\geq  \frac{1}{4} \sum \limits_{u_1,u_2} H(K^{u_1,u_2}_T)- \text{leak}^{EC}_{obs} - h(e)&
\end{flalign}
where $K^{u_1,u_2}_T$ denotes the key data of the node which defines the protocol, and $e$ is the overall error rate between Alice's raw key $d_A$ and Bob's corrected raw key $d_B^\prime$.


\section{Realistic Considerations}\label{s:decoy}

The key rates found in the previous sections are constructed for the ideal case where the legitimate parties exchange qubits over a lossless channel, and Eve performs collective attacks. Most QKD implementations use optical modes to transmit signal states over a lossy channel. This is problematic, as Eve can actively select for pulses that contain additional information. Additionally, Eve is not limited to collective attacks. In this section, we will address these concerns. 


\subsection{From Collective Attacks to General Attacks}\label{ss:catga}

First, let us justify our examination of collective attacks in the previous sections. The techniques developed in Ref. \cite{Christandl09} provide a method for extending the security of collective attacks to general attacks, provided the protocol is invariant with respect to permutations of the input states. Even when exchanging qubit signal states in a lossless setting, the STR protocol is \emph{not} permutation invariant, due to the fact that basis sifting occurs before the parity announcement. 

We may still extend the security of the STR protocol to general attacks by noting that key elements of the STR protocol are permutation invariant. The process of basis sifting commutes with the measurement process; therefore, basis sifting may be viewed as an initial step preceding measurements. In a general attack, Eve may store all the signal states in a large quantum memory before distributing them to the legitimate parties. Let us denote this state with $\rho^N$. The initial step of basis sifting maps the state $\rho^N$ to a smaller state $\rho^n_\text{data}$.
Importantly, the remainder of the STR protocol acting on $\rho^n_\text{data}$ (measurements and post-processing) is permutation invariant. This assumes privacy amplification is carried out in a permutation invariant manner. While there exist permutation invariant methods of privacy amplification, not all methods of privacy amplification fit this criteria.

In order to separate the initial sifting step from the remainder of the STR protocol, a common proof technique is to assume Eve performs the sifting step. However, Eve is unable to perform the process of basis sifting as she does not have access to the basis information. We therefore posit a hypothetical protocol that proceeds identically to the STR protocol; however, before the protocol begins, whether signals will be kept or discarded during sifting is announced to Eve for each time slot (but not the basis information). Without loss of generality, we may now assume Eve removes any signals that would be lost in basis sifting. The steps following sifting are now permutation invariant with respect to the signals grouped by the parity announcements. We may now assume that Eve performs collective attacks on these signals, when considering the infinite key limit \cite{Christandl09}. This is precisely the situation we have analyzed in previous sections.

Importantly, the key rate for this hypothetical protocol is a lower bound on the STR protocol. Therefore the analysis given in the previous sections provides a lower bound on the key rate of the STR protocol, when considering general attacks on qubit signal states sent over lossless channels.

We may extend the above analysis to include qubit signals exchanged over lossy channels. In the case of a lossy channel, the legitimate parties discard any data in which a detector did not click. Similar to basis sifting, the removal of these events commutes with the measurement process. Despite being under Eve's control, we may treat vacuum sifting in the same manner as basis sifting. The remainder of the argument follows similarly.


\subsection{STR Protocol with Decoy States}\label{ss:decoy}

Recall that when considering qubit-level security, our analysis directly applied to a number of protocols with different variations of state creation and measurement (see the discussion ending Section \ref{s:proto}). In order to move beyond qubit-level security, we restrict our focus to the BB84 STR protocol as outlined in the beginning of Section \ref{s:proto}. While the following analysis still applies to the case where the roles of state preparation and measurement are interchanged, it is not straightforward to generalize the analysis to entanglement-based or measurement-device independent STR protocols.

Current QKD implementations do not have access to ideal single photon sources. Typically, highly attenuated lasers are used to generate the signal states. These sources are described by coherent states, where the photon number adheres to a Poisson distribution. The probability of sending multi-photon pulses is therefore non-zero. An eavesdropper can exploit multi-photon pulses through a \emph{photon number splitting attack} \cite{Luteknhaus00,Luteknhaus02}. In order to improve the key rate for realistic sources, the legitimate parties may employ \emph{decoy state analysis} \cite{Hwang03, Lo05, Wang05}, supported by \emph{tagging} \cite{Gottesman04,Inamori07}. In addition to the original pulses, the legitimate parties send decoy states which have a variable mean photon number, $\mu _n$. By introducing these additional observables, an upper bound may be estimated on the number multi-photon events. On the detection side, squashing methods can deal with the possibility of multiple photons entering a detector \cite{Beaudry08,Tsurumaru08,Moroder10}. Given the existence of a squashing map, the detection pattern can be interpreted as if it resulted from a vacuum or single-photon pulse.

Due to the structure of the parity announcement, if $\emph{any}$ of the legitimate parties emits a multi-photon pulse, Eve may perform a photon number splitting attack. For simplicity, we assume Eve obtains full information of the corresponding raw key bit whenever this happens --- except when a vacuum signal is sent in the first link. (Due to dark counts, a detector may still click, even when a vacuum signal was sent.) Additionally, we need to rescale privacy amplification to account for Eve's interaction with the single photon pulses. Let us denote the fraction of detected events used for the raw key in which  a single photon was sent in the first link and all other links sent a single photon (vacuum or single photon) pulse to be $f_{s,s}$ ($f_{s,v/s}$). Similarly, $ f_{v}$ is the fraction of events where a vacuum pulse was sent in the first link, and $e_{s,s/v}^{\textbf{u}}$ is the error rate arising from events where a single photon was sent in the first link and all other links sent vacuum or single photon pulses. The fraction of multi-photon events is then given by $f_m =1 - f_{v} - f_{s,v/s}$. This is directly subtracted from the key rate. For $m$ nodes, the corresponding decoy state key rate is 
\begin{flalign}\label{decoy}
r \geq& \sum_{\substack{u_1, \ldots ,u_{m+1}  }}p_{\textbf{u}} H(K^{\textbf{u}}_A)-\text{leak}^{EC}_{obs} \\
&-f_{s,s/v}\left( \sum_{\substack{u_1, \ldots ,u_{m+1} }} p_{\tilde{\textbf{u}}} \    \ h(e_{s,s/v}^{\textbf{u}})\right)  - (1 - f_{v} - f_{s,v/s}) & \nonumber
\end{flalign}
for the asymptotic limit. Again, $\textbf{u} := \{ u_i\}$ represents the basis choice in each link, and $\tilde{\textbf{u}} :=  \{ u_i\oplus 1\}.$ Note that in practice, the fraction of events in which a single photon was sent in the first link and all other links sent a vacuum ($f_{s,v}$) will be small. The approximations $f_{s,s/v} \approx f_{s,s} $ and $e_{s,s/v} \approx e_{s,s}$ will safely lower bound the key rate.

Decoy state analysis has been thoroughly explored in the literature. Most techniques assume that the detected signal states are independent and identically distributing (i.i.d.). In general, this assumption is not valid. For this reason, we rely on the analysis found in Ref. \cite{Curty13}, which does not assume  i.i.d.~signal states. The analysis in Ref. \cite{Curty13} uses observables arising from $\rho^N$  (the overall state shared between the legitimate parties, including vacuum and multi-photon signals) to bound the fraction of tagged signals and the single-photon error rate by use of decoy states. While the analysis directly applies for a single node, it may be extended to an STR protocol with an arbitrary number of nodes. Observations from $\rho^N$ provide a promise about the fraction of detected signals that are tagged. The same analysis from Section \ref{ss:catga} may be applied to extend the security analysis from collective attacks to general attacks; the final key rate simply needs to be updated with respect to this promise, as shown in Eq (\ref{decoy}).

 \begin{figure}
\begin{center}
\includegraphics[scale=0.8]{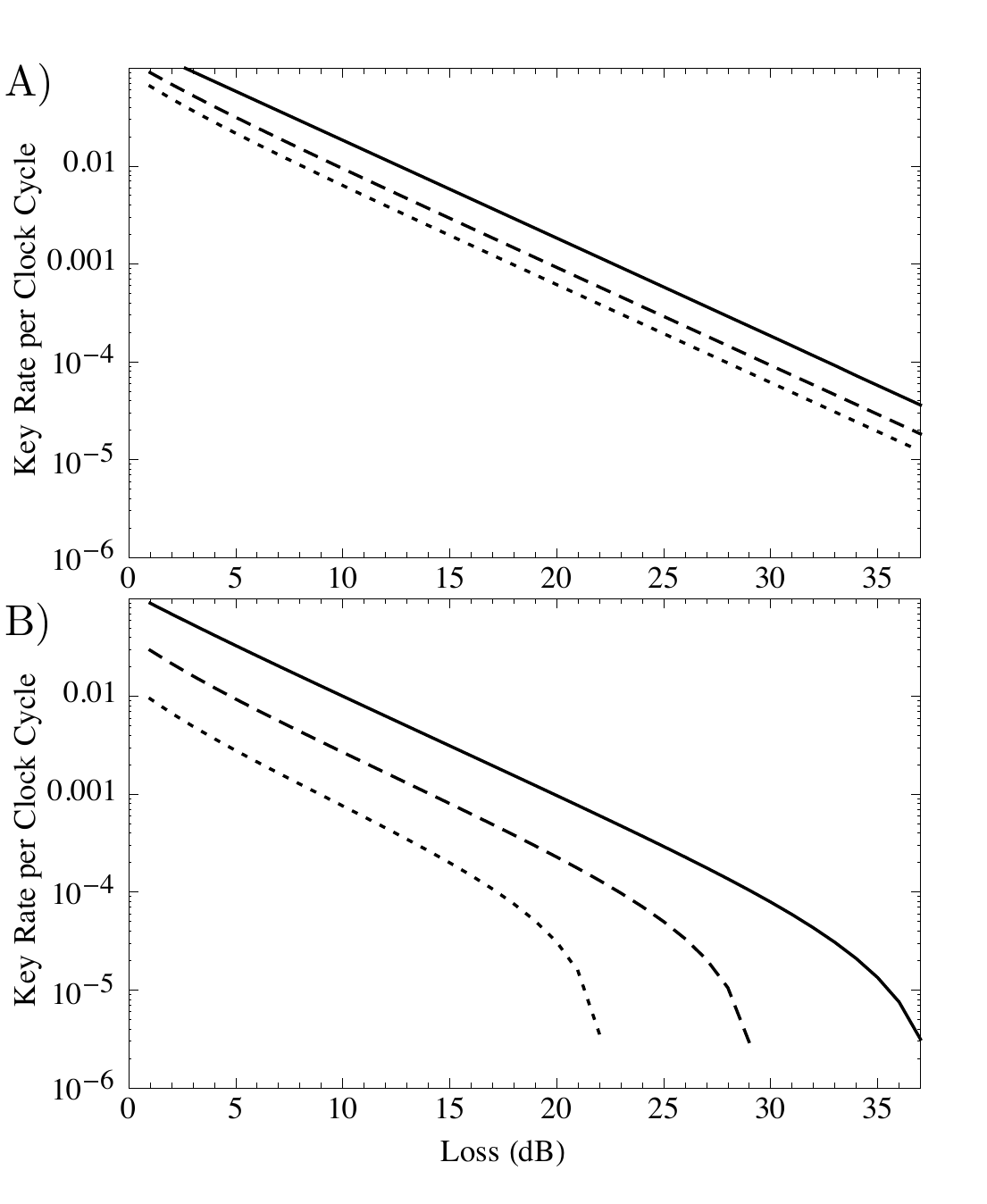} 
\caption{The number of secret key bits generated per clock cycle, for a conventional trusted node (solid line), an STR with one node (dashed line), and an STR with two nodes (dotted line), as a function of the loss in a single link. (A) The scaling is examined in the error-free limit. (B) We use an intrinsic error rate of 1.85\% and a dark count rate of $6\times 10^{-6}$ per clock cycle. Furthermore, we assume the information lost during error correction is $1.2$ times greater than the Shannon limit. For both cases, we assume a detector efficiency of 50\%. For each value of loss, the signal intensity is optimized. Finite size effects are not considered, and Alice and Bob are assumed to perfectly determine $f_{v}$ and $f_{s,v/s}$.
}
\label{fig_rate_distance}
\end{center}
\end{figure}

In Fig. \ref{fig_rate_distance}, we compare the STR protocol to a conventional trusted relay using decoy state techniques. We plot the key generation rate per clock cycle as a function of channel loss in one arm of the relay, optimizing the mean photon intensity, $\mu$, at each distance. Though the key generation rate per clock cycle is lower for an STR, the computational load on each node is also reduced. In situations where the nodes have limited computational power, the key generation rate \emph{per unit time} may in fact be greater for an STR than a conventional trusted relay.


\section{Conclusion}
We have examined the security of a simplified trusted relay which implements the quantum phase of the BB84 protocol. In addition to providing an analytic key rate for an ideal case (lossless and i.i.d.~qubit signals), we have provided a clear path for extending the proof to realistic circumstances (loss, general attacks, and weak coherent states). Our results show that a trusted relay can function securely without the nodes participating in parameter estimation, error correction or privacy amplification. 

In comparison to conventional trusted relays, the STR protocol benefits from its simplicity; however, this comes at the cost of a lower key rate. Compounding errors prevent a naive implementation of the STR protocol from achieving arbitrarily long distances; however, several STRs may be chained together (similar to a conventional trusted relay) to form a pattern of simplified and conventional trusted relay nodes. For many scenarios, the optimal network likely incorporates STRs and conventional trusted relays. In addition, for situations where the intermediate nodes are limited in computational power or communication bandwidth (such as lightweight satellites), STRs may prove to have a higher key generation rate per second, as they reduce the computation and communication requirements for the intermediate nodes.


\appendix

\section{Invariance of $\rho_{ATBE}$ with respect to Pauli-operations}\label{appendix1}

In this appendix, we explicitly show that states derived by applying correlated Pauli operations to $\rho_{ATBE}$ result in the same basis-dependent error rate and Holevo quantity. For clarity, let us change our original notation from $\rho^{i}_{ATBE}$ to $\rho^{U_{r,s}U_{r^\prime,s^\prime}}_{ATBE}$. The set of states $\beta = \{\rho^{U_{r,s}U_{r^\prime,s^\prime}}_{ATBE}\}$ is given explicitly by
\begin{flalign}
\rho^{U_{r,s}U_{r^\prime,s^\prime}}_{ATBE}& := U_{r,s} \otimes U_{r,s} \otimes U_{r^\prime,s^\prime} \otimes U_{r^\prime,s^\prime}  \otimes \id_E \rho_{ATBE}&\\
 &(U_{r,s} \otimes U_{r,s} \otimes U_{r^\prime,s^\prime} \otimes U_{r^\prime,s^\prime}\otimes\id_E)^\dagger. &\nonumber
\end{flalign}
For our purposes, we define the Pauli matrices as
\begin{equation}
U_{r,s} := \sum \limits_{k=0}^{1} (-1)^{ks} |k+r\ket\bra k |
\end{equation}
for $r,s \in \{0,1\}$.

To show that the basis-dependent error rates are invariant for each state $\rho^{U_{r,s}U_{r^\prime,s^\prime}}_{ATBE}$, we make use of the fact that Pauli matrices only permute the BB84 signal states \emph{within} each basis. Let us define the action of the Pauli operator $U_{r,s}$ on the signal state $|\phi^u_x\ket$ to be $U_{r,s} |\phi^u_x\ket = |\phi^u_{x\oplus h(u,r,s)}\ket$ for some function $h(u,r,s)$ with binary output. It follows that this relation similarly applies to the BB84 POVM elements. 
The averaged error rate arising from the state $\rho^{U_{r,s}U_{r^\prime,s^\prime}}_{ATBE}$ can then be rewritten as
\begin{flalign}
e^{u_1,u_2}_{U_{r,s}U_{r^\prime,s^\prime}}&= \sum \limits_{b} \sum \limits_{x,t }tr_{AB}(M^{u_1}_{A,x}\otimes M^{u_1}_{T,t} \otimes&\\
& M^{u_2}_{T,t\oplus b} \otimes M^{u_2}_{B,x\oplus b\oplus1} \rho^{U_{r,s}U_{r^\prime,s^\prime}}_{ATB})&\nonumber\\
&= \sum \limits_{b} \sum \limits_{x,t}tr_{AB}(M^{u_1}_{A,x\oplus h(u_1,r,s)}\otimes M^{u_1}_{T,t\oplus h(u_1,r,s)}&\nonumber\\
& \otimes M^{u_2}_{T,t\oplus b\oplus h(u_2,r^\prime,s^\prime)} \otimes M^{u_2}_{B,x\oplus b\oplus h(u_2,r^\prime,s^\prime)\oplus1} \rho_{ATB}).& \nonumber
\end{flalign}
Let us define $x^\prime := x\oplus h(u_1,r,s), t^\prime := t\oplus h(u_1,r,s)$, and $b^\prime := b\oplus  h(u_1,r,s) \oplus  h(u_2,r^\prime,s^\prime)$. It follows
\begin{align}
e^{u_1,u_2}_{U_{r,s}U_{r^\prime,s^\prime}}=& \sum \limits_{b^\prime} \sum \limits_{x^\prime,t^\prime }tr_{AB}(M^{u_1}_{A,x^\prime}\otimes M^{u_1}_{T,t^\prime} \otimes&\\
& M^{u_2}_{T,t^\prime \oplus b^\prime} \otimes M^{u_2}_{B,x^\prime \oplus b^\prime \oplus1} \rho_{ATB})&\nonumber \\
=&\ e^{u_1,u_2}&
\end{align}
Therefore the basis-dependent error rates are invariant when the same Pauli-operation is applied in each link.

In order to show that the Holevo quantity is invariant, we need examine the probability of obtaining a key bit ($p_k$) and Eve's conditional states ($\rho_E^{k}$). Given that the trace is invariant under a global unitary, the probability distribution $p_k$ is simply permuted: 
\begin{flalign}
p^{U_{r,s}U_{r^\prime,s^\prime}} _k &:=tr_{ATBE} \{  M^{u_1}_{A,k} \otimes \id_{TBE} \rho^{U_{r,s}U_{r^\prime,s^\prime}}_{ATBE}  \} = p_{k^\prime}.&
\end{flalign}

It follows similarly that the action of the local Pauli operations simply permutes Eve's conditional state,
$\rho_E^{k,U_{r,s}U_{r^\prime,s^\prime}}  =\rho_E^{k^\prime}.$
Therefore by expressing the Holevo quantity in terms of the von Neumann entropy, it follows that
\begin{flalign}
\chi&\left(\sum_b p_{b|u_1,u_2} \rho^{u_1,u_2,b,U_{r,s}U_{r^\prime,s^\prime}}_{ATBE} \otimes |b\ket \bra b|_C,\textbf{M}^{u_1}_A\right)  =& \nonumber \\
&\ \chi\left(\sum_b p_{b|u_1,u_2} \rho_{ATBE}^{u_1,u_2,b} \otimes |b\ket \bra b|_C,\textbf{M}^{u_1}_A\right). &
\end{flalign}


\section{Form of symmetrized state $\rho^\text{Bell}_{ATBE}$}\label{appendix2}

The form of $\rho^{\text{Bell}}_{ATBE}$ is calculated using an extension of the work presented in Ref. \cite{FerencziThesis}. 
We can express the reduced average state as:
\begin{flalign}
\frac{1}{16} \sum\limits_{r,s,r^\prime,s^\prime} \rho_{ATB}^{U_{r,s}U_{r^\prime,s^\prime}}  &=&\\
 \frac{1}{16}\sum \limits_{r,s,r^\prime,s^\prime}& U_{r,s} \otimes U_{r,s} \otimes U_{r^\prime,s^\prime} \otimes U_{r^\prime,s^\prime}& \nonumber\\
 &\rho_{ATB} (U_{r,s} \otimes U_{r,s} \otimes U_{r^\prime,s^\prime}  \otimes U _{r^\prime,s^\prime} )^\dagger .&\nonumber
\end{flalign}
 Any state $\rho_{ATB}$ can be expressed in the tensored Bell basis $\mathcal{B} = \{|U_{r,s}\ket \otimes |U_{r^\prime,s^\prime}\ket : r,s,r^\prime,s^\prime=0,1\}$. Note that the action of the Pauli matrices on a basis element is
\be
U_{r,s} \otimes U_{r,s} \otimes U_{r^\prime,s^\prime} \otimes U_{r^\prime,s^\prime}|U_{m,n}\ket \otimes |U_{p,q}\ket\\
 = (-1)^{-sm+rn-s^\prime p +r^\prime q}|U_{m,n}\ket \otimes |U_{p,q}\ket. \nonumber
\ee 
By averaging over the Pauli matrices, we find
\begin{flalign} \label{A4}
\sum \limits_{r,s,r^\prime,s^\prime} & U_{r,s} \otimes U_{r,s} \otimes U_{r^\prime,s^\prime} \otimes U_{r^\prime,s^\prime} |U_{m,n}\ket \bra U_{m^\prime,n^\prime}|
\otimes &\nonumber \\
 &|U_{p,q}\ket\bra U_{p^\prime,q^\prime}| (U^*_{r,s} \otimes U_{r,s} \otimes U_{r^\prime,s^\prime}  \otimes U^* _{r^\prime,s^\prime} )^\dagger \nonumber&\\
=& \sum \limits_{r,s,r^\prime,s^\prime} (-1)^{-s(m+m^\prime)+r(n+n^\prime)-s^\prime(p+p^\prime)+r^\prime(q+q^\prime)}& \nonumber \\
&|U_{m,n}\ket \bra U_{m^\prime,n^\prime}|  \otimes |U_{p,q}\ket\bra U_{p^\prime,q^\prime}|. &
\end{flalign}
The off diagonal elements can be shown to vanish by observing that the coefficient in Eqn. \ref{A4} is equivalent to $16\delta_{m,m^\prime}\delta_{n,n^\prime}\delta_{p,p^\prime}\delta_{q,q^\prime}$. Therefore the averaged state can be expressed simply as 
\begin{flalign}
\frac{1}{16} \sum\limits_{r,s,r^\prime,s^\prime} \rho_{ATB}^{U_{r,s}U_{r^\prime,s^\prime}} &=\nonumber\\
 \sum \limits_{m,n,p,q}  \alpha_{m,n,p,q} &|U_{m,n}\ket \bra U_{m,n}|  \otimes  |U_{p,q}\ket \bra U_{p,q}| .&
\end{flalign}

\bibliography{STNbib}{}
\bibliographystyle{nature}

\end{document}